\documentclass[conference]{IEEEtran}

\ifCLASSINFOpdf
  \usepackage[pdftex]{graphicx}
\else
  \usepackage[dvips]{graphicx}
\fi
\usepackage[cmex10]{amsmath}
\usepackage[tight,footnotesize]{subfigure}
\usepackage{stfloats}

\begin{document}

\title{Network Coded Transmission of Fountain Codes over Cooperative Relay Networks}

\author{\IEEEauthorblockN{E. Kurniawan, S. Sun, K. Yen, and K. F. E. Chong}
\IEEEauthorblockA{Institute for Infocomm Research\\
Agency for Science Technology and Research\\
1 Fusionopolis Way, \#21-01 Connexis (South Tower), Singapore 138632\\
Email: \{ekurniawan,sunsm,yenkai,kfchong\}@i2r.a-star.edu.sg}}

\maketitle

\begin{abstract}
In this paper, a transmission strategy of fountain codes over cooperative relay networks is proposed. When more than one relay nodes are available, we apply network coding to fountain-coded packets. By doing this, partial information is made available to the destination node about the upcoming message block. It is therefore able to reduce the required number of transmissions over erasure channels, hence increasing the effective throughput. Its application to wireless channels with Rayleigh fading and AWGN noise is also analysed, whereby the role of analogue network coding and optimal weight selection is demonstrated.
\end{abstract}

\IEEEpeerreviewmaketitle

\section{Introduction}
Fountain code \cite{Fountain:MacKay} and cooperative communication \cite{User:Sendonaris} are two transmission strategies which are gaining popularity in recent years. In fountain codes, message bits are grouped into blocks, each containing several packets. Encoding is performed by taking random linear combination of the packets within each block over a Galois Field (typically $GF(2)$). As opposed to other fixed rate scheme, with fountain code, the source continuously transmits encoded packets until positive acknowledgement is received. Hence, its optimality is guaranteed for erasure channels, regardless of the erasure probability.

Cooperation, on the other hand, improves transmission quality by making use of neighbouring nodes to forward the message to destination. By creating multiple paths between source and destination (each subjected to independent fading), a diversity advantage can be exploited. Application of fountain codes in cooperative network have also been studied. For example, reference \cite{Rateless:Castura} and \cite{Raptor:Zhang} proposed to combine distributed space time block coding (DSTBC) and fountain code for single carrier and multiple carrier transmission respectively, and showed that extra diversity gain can be achieved. In \cite{TheDesign:Puducheri}, direct application of fountain code in cooperative network is analysed. It was shown that careful degree distribution design is necessary to ensure good decoding performance. Alternatively, using conventional degree distribution, the encoding/decoding process at the relay node can be modified to cater for the online re-coding requirement, as discussed in \cite{Relaying:Gummadi}.

Although it has been shown that performance improvement can be achieved using fountain codes in cooperative networks \cite{Performance:Molisch}, this advantage brings about extra complexity, especially when more than one relay node is involved. Motivated to address this issue, recently the authors have proposed an amplitude modulation scheme for fountain code transmission over multiple relay cooperative networks \cite{Transmission:Kurniawan}. However, the scheme is developed for erasure channels, and it is not directly applicable to wireless fading channels. The focus of this paper is to find an alternative strategy to tackle the above issue. Here, a novel scheme that combines network coding \cite{Network:Ahlswede} and fountain codes transmission is proposed. In erasure channel, the scheme applies digital network coding onto fountain encoded packets of two consecutive blocks, and allows source node to transmit together with the successful relay. Whereas in wireless channel, the scheme employs analogue network coding with appropriate power allocation. The performance of the scheme is then analysed numerically, and it is shown to improve the overall throughput in both types of channel.

The rest of this paper is organised as follows. Section \ref{Sec:SystemModel} describes the system model. The proposed transmission scheme and its application into wireless channel are given in \ref{Sec:TransScheme} and \ref{Sec:AppsToWireless} respectively. Numerical results are then presented in Section \ref{Sec:Numerical}. Finally, Section \ref{Sec:Conclusion} gives concluding remarks.

\begin{figure}[t]
	\centering
		\includegraphics[width=200pt]{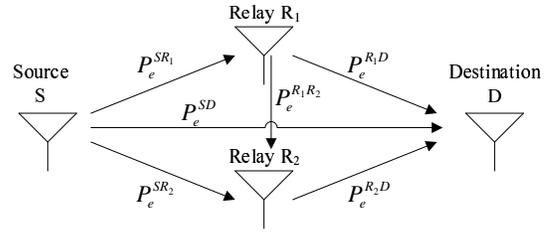}
	\caption{System diagram of two-relay cooperative erasure network}
	\label{fig:ErasureNetwork}
\end{figure}

\section{System Model}
\label{Sec:SystemModel}
A half duplex cooperative network with one source $S$, one destination $D$, and two relay nodes (denoted as $R_1$ and $R_2$) as depicted in Figure \ref{fig:ErasureNetwork} is considered\footnote{Although the discussion presented in this paper is mainly for two relay nodes scenario, the results can be extended into general number of relays.}. At any one time, a node can either transmit or receive, but not both simultaneously. The channel between any given two nodes is modelled as erasure channel, with a superscript indicating the nodes under consideration (e.g., $P_e^{SD}$ is used to indicate the erasure probability of the channel between $S$ and $D$).

Message bits at the source are grouped into blocks of $K$ packets, each composed of $m$ information bits. Fountain code is then applied to the $K$ packets in which linear combination of randomly selected $d$ (generated following some degree distribution $\rho(d)$) out of $K$ packets are transmitted. For simplicity, random linear fountain code is considered throughout the analysis. Therefore, the degree distribution used to generate encoded packets can be expressed as:
\begin{equation}
\label{eqn:RhoD}
\rho(d) =
\begin{cases}
0 		& \text{for $d=0$}
\\
C_d^K / (2^{K}-1) & \text{for $0<d\leq K$}
\end{cases}
\end{equation}
which can be approximated as $\rho(d) \approx C_d^K / 2^{K}$ for large $K$. Here, $C_n^m$ denotes the number of combinations for selecting $n$ out of $m$ elements. The analysis presented in this paper is independent of the actual degree distribution used, therefore the same technique can be applied to other (more practical) degree distributions such as Robust Soliton Distribution \cite{LT:Luby}.

\section{Transmission Schemes}
\label{Sec:TransScheme}

\subsection{Direct Transmission (Without Relay)}
As a baseline comparison, direct transmission of fountain code from $S$ to $D$ is considered. In this case, the number of encoded packets received unerased at $D$ (denoted as $N$) is random, and it is related to the number of transmitted packets ($M$, where $0 \! \leq \! N \! \leq \! M$) through binomial distribution with parameter $P_e^{SD}$ as follows:
\begin{equation}
\label{eqn:Binom}
\mathcal{B}_{M, P_e^{SD}}\left(N \right) = C_N^M \: (1-P_e^{SD})^N \: 	(P_e^{SD})^{M-N} \
\end{equation}
Given that $N$ encoded packets are available at $D$, the probability that the corresponding $K \times N$ generator matrix is full rank can be calculated. Following the assumption that random linear fountain code is used, all $2^{KN}$ possible binary generator matrices are equiprobable\footnote{Note that we have included the probability of generating degree 0 column to simplify the expression. This approximation is tight for large $K$.}. Hence, the probability of successful decoding can be calculated as:
\begin{eqnarray}
\label{eqn:TheCDF}
F(N) &=& \begin{cases}
0 & \text{$N < K$}\\
\prod_{i=0}^{K-1}\left( 1 - 2^{i-N} \right) & \text{$N \geq K$}
\end{cases}
\end{eqnarray}
The above equation can be interpreted as the cumulative distribution function (CDF) of successful decoding after $N$ packets are received. The respective probability distribution function (PDF) can be expressed as:
\begin{eqnarray}
\label{eqn:pm}
f(N) \hspace{-7pt} &= \hspace{-7pt}& F(N) - F(N-1) \nonumber \\
\hspace{-7pt}&= \hspace{-7pt}& \begin{cases}
0 & \text{$N < K$}\\
\prod_{i=0}^{K-1}\left( 1 - 2^{i-N} \right) & \text{$N = K$}\\
\frac{2^{-N} (2^K - 1)}{1 - 2^{K-N}} \: \prod_{i=0}^{K-1}\left( 1 - 2^{i-N+1} \right) & \text{$N > K$}
\end{cases}
\end{eqnarray}
Noting that the transmission stops as soon as $D$ is able to decode, the PDF of successful decoding after $M$ transmissions is given as:
\begin{equation}
\label{eqn:p1}
p^{(1)}(M, P_e^{SD}) = \frac{1-P_e^{SD}}{\Omega} \! \sum_{i=K-1}^{M-1} \!\! \mathcal{B}_{M - 1, P_e^{SD}}\left(i \right) f(i+1)
\end{equation}
where $\Omega$ is the normalising constant to satisfy the unit-sum constraint $\sum_M p^{(1)}(M, P_e^{SD})=1$, while $\mathcal{B}(\cdot)$ and $f(\cdot)$ are given in (\ref{eqn:Binom}) and (\ref{eqn:pm}) respectively.

\subsection{Naive Relaying Scheme}
\begin{figure}[t]
  \begin{center}
    \mbox{
      \subfigure[Phase 1]{\includegraphics[width=110pt]{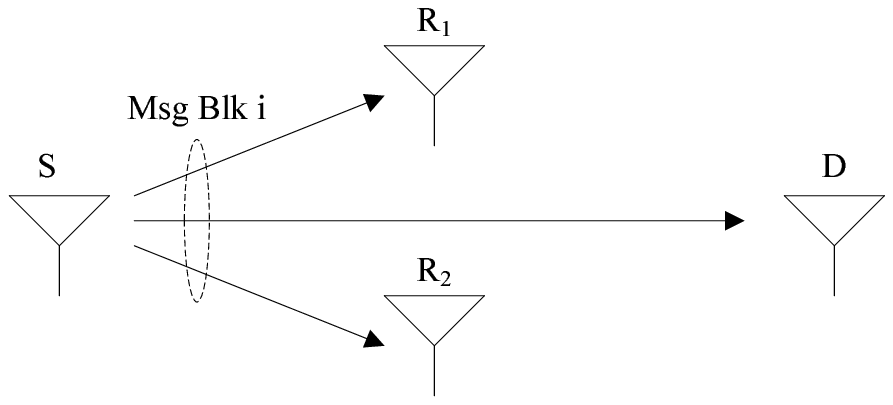}} \quad
      \subfigure[Phase 2]{\includegraphics[width=110pt]{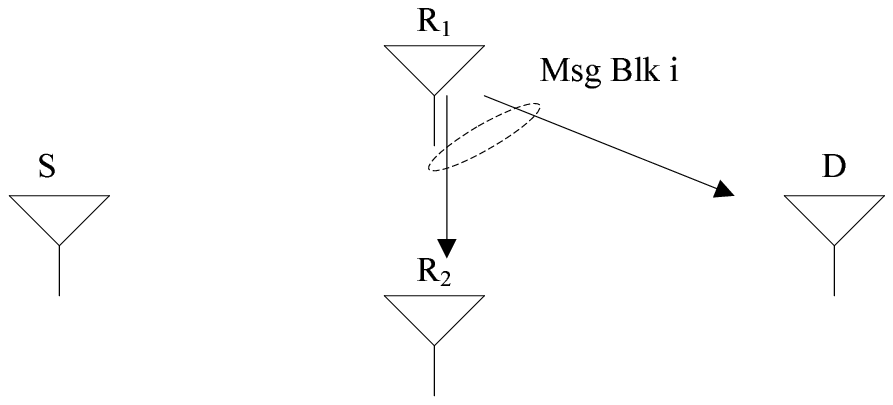}}
      }
    \caption{Illustration of two-phase transmission using naive relaying scheme (assuming $R_1$ is the successful relay).}
    \label{fig:ConventionalRelay}
  \end{center}
\end{figure}

In the presence of relay node, a natural way to perform transmission is through multiple hops using two-phase scheme as illustrated in Figure \ref{fig:ConventionalRelay}. During the first phase, $S$ broadcasts fountain encoded packets to all $R$'s, as well as $D$. Since $P_e^{SR}$ (assumed to be equal for all relays) is smaller than $P_e^{SD}$ due to geographical proximity, it is highly likely that one of the $R$'s is able to decode the message before $D$. Denoting the total number of relays as $\mathcal{R}$, the probability that any of them is able to decode after $j$ transmissions can be calculated as:
\begin{equation}
\label{eqn:RelayDecode}
q(j) = \sum_{r=1}^{\mathcal{R}} C_r^{\mathcal{R}} \left[p^{(1)}(j, P_e^{SR})\right]^r [\sum_{i>j} p^{(1)}(i, P_e^{SR}) ]^{\mathcal{R}-r}
\end{equation}
The above equation holds for any value of $\mathcal{R}$, and it determines the duration of phase one transmission. Once any of the $R$'s decodes, the transmission enters the second phase, whereby the successful $R$ transmits fountain encoded packets to $D$. Since the last packet from $R$ to $D$ must not be erased, the PDF of successful decoding after $M$ transmissions is given as:
\begin{eqnarray}
\label{eqn:p2}
&& \hspace{-20pt} p^{(2)}(M) = p^{(1)}(M, P_e^{SD}) [\sum_{j \geq M} p^{(1)}(j, P_e^{SR})]^\mathcal{R} + (1-P_e^{RD}) \nonumber \\
&& \hspace{-20pt} \sum_{j<M} \! q(j) \! \left(\sum_{s=0}^j  \!\! \sum_{t=0}^{M-j-1} \!\!\! \mathcal{B}_{j, P_e^{SD}}(s) \mathcal{B}_{M-j-1, P_e^{RD}}(t) f(s\!+\!t\!+\!1) \!\! \right)
\end{eqnarray}
where the first term represents the case when $D$ is able to decode before any of the $R$'s, hence the probability of successful decoding falls back to direct transmission case; and the second term represents the case when one of the $R$'s is able to decode before $D$, hence the received packets come from both $S$ and $R$ during the first and second phase respectively.

\textbf{\textit{Remark:}} Due to broadcast nature of the transmission, the other relays can receive the packets from the transmitting relay during phase two. However, those packets would not be useful, as they only contain information about the current message block. The moment $D$ is able to decode, it will send positive acknowledgement (ACK), and a new protocol cycle starts. Here, ideal ACK channel is assumed available for simplicity.

\subsection{Proposed Network Coded Scheme}
\begin{figure}[t]
  \begin{center}
    \mbox{
      \subfigure[Phase 1]{\includegraphics[width=110pt]{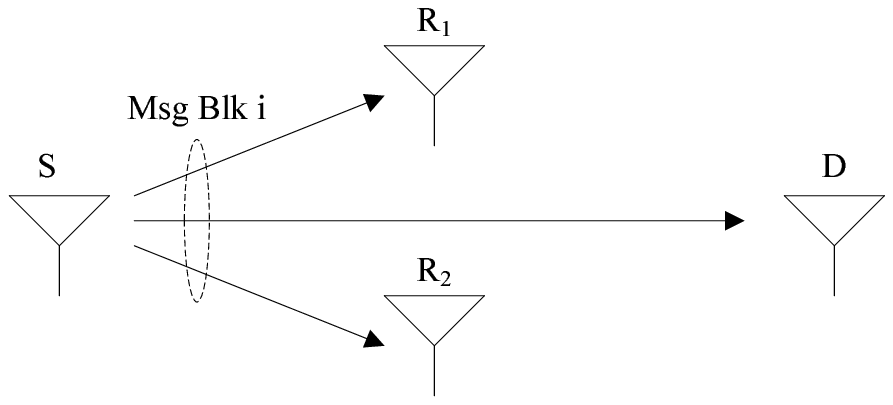}} \quad
      \subfigure[Phase 2]{\includegraphics[width=110pt]{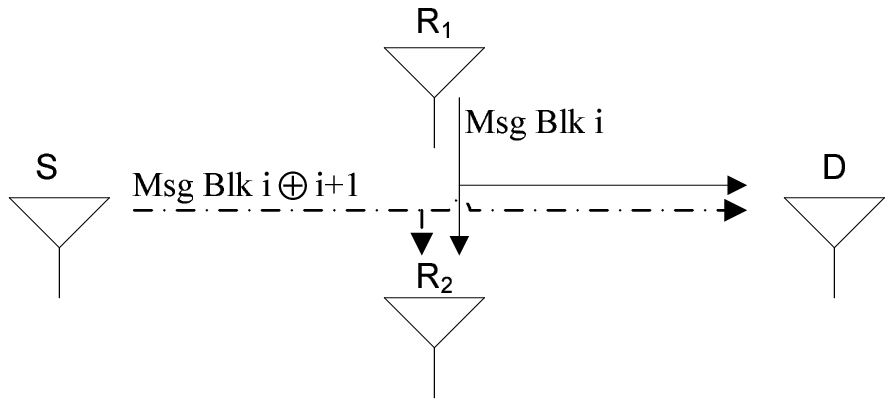}}
      }
    \caption{Illustration of two-phase transmission using the proposed network coded scheme (assuming $R_1$ is the successful relay).}
    \label{fig:ProposedScheme}
  \end{center}
\end{figure}

The idea of this scheme is to allow $S$ transmitting partial information about the next message block during the second phase, hence enabling the receiving relays to help the transmission of subsequent message blocks. The first phase transmission in this case is identical to the naive relaying scheme. On the second phase, in addition to the successful $R$, $S$ also transmits the network coded version of fountain codes for the current and the next message block as illustrated in Figure \ref{fig:ProposedScheme}. Here, the fountain code for the current message block at the successful $R$ and $S$ are identical, while different code can be used for the next message block at $S$.

Without loss of generality, let $R_1$ be the successful relay, and there are only two relays in the network. Since the channel is an ideal erasure channel, the effective packet received at $R_2$ and $D$ is the combination (bit-wise exclusive-or) of the packets transmitted from $S$ and $R_1$. Depending on which of those packets are erased, $D$ would store the resulting packet to one of the three buffers it maintains as illustrated in Table \ref{tab:Buffer}. When the packet from $R$ is the only one unerased, the received packet contains information about the current block, and it is stored in buffer $1$. When the packet from $S$ is the only one unerased, the packet contains a network coded version of the current and next block, and it is stored in buffer $2$. Finally, when both packets are unerased, the received packet contains information about the next block, and it is stored in buffer $3$.
\begin{table}[t]
	\centering
		\caption{Buffer assignment table at destination node during phase two.}
		\label{tab:Buffer}
		\begin{tabular}{c}
					\includegraphics[width=228pt]{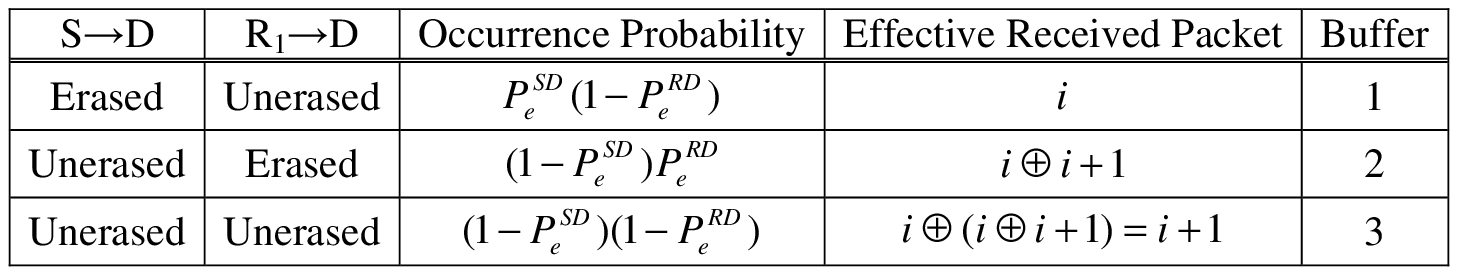}
		\end{tabular}
\end{table}

\begin{figure}[!h]
	\centering
		\includegraphics[width=136pt]{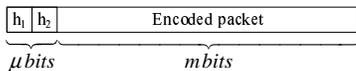}
	\caption{Modified packet structure with multiple headers.}
	\label{fig:PacketStructure}
\end{figure}

To enable the receiving nodes to identify which one of the three cases has happened, a modification is made to the packet structure, as depicted in Figure \ref{fig:PacketStructure}.
\\By assigning header $h_1$ exclusively to $S$ and $h_2$ to $R$, the first $\mu$ bits of the packet can be used to identify which packets has been erased. Meanwhile, the remaining $m$ bits are shared by both $S$ and $R$. As long as $m >> \mu$, the overhead caused by the additional header information is negligible.

Comparing this scheme with the naive relaying, several observations can be made. Firstly, the useful fraction of transmissions during the second phase can be calculated by adding the probabilities of the three cases, which is equal to $(1\!\!-\!P_e^{RD})\!+\!(1\!\!-\!P_e^{SD})P_e^{RD}$. This is larger than that of the naive relaying method, which is only $(1\!-\!P_e^{RD})$. Secondly, the packets received at $R_2$ during phase two contain information about the next message block, which can accelerate its transmission to $D$ once the current message is decoded.

To perform the combinatorial analysis of this scheme, the following auxiliary function is defined:
\begin{eqnarray}
\hspace{-8pt}g_{n_p, P_e}(M) \hspace{-9pt} &= \hspace{-9pt}& \begin{cases}
\hspace{-2pt}\frac{f_{n_p}(M)}{\Omega_2} & \hspace{-7pt} \text{$M = 0$}\\
\hspace{-2pt}\frac{1-P_e}{\Omega_2} \sum_{i=0}^{M-1} \beta_{M-1, P_e}(i) \: f_{n_p}(i+1) & \hspace{-6pt} \text{$M > 0$}\\
\end{cases}
\end{eqnarray}
where $\Omega_2$ is the normalisation factor required to make $\sum_M g_{n_p, P_e}(M) = 1$. Function $f_{n_p}(.)$ is defined as:
\begin{eqnarray}
f_{n_p}(N) \hspace{-7pt} &= \hspace{-7pt}& \begin{cases}
F(n_p) & \text{$N = 0$}\\
F(N + n_p) - F(N + n_p - 1) & \text{$N > 0$}\\
\end{cases}
\end{eqnarray}
which can be evaluated by substitution with equation (\ref{eqn:TheCDF}). The above auxiliary function $g_{n_p, P_e}(M)$ can be interpreted as the probability of successful decoding (in point-to-point case) after $M$ transmissions, given that $D$ already has $n_p$ packets available. Denote the number of packets available at $R_1$, $R_2$, and $D$ as $n_1$, $n_2$, and $n_3$ respectively. The probability that either $R_1$ or $R_2$ is able to decode after $j$ transmissions is:
\begin{equation}
\tilde{q}(j) \! = \! g_{n_1, P_e^{SR}}(j) \! \sum_{k>j} g_{n_2, P_e^{SR}}(k) + g_{n_2, P_e^{SR}}(j) \! \sum_{k\geq j} g_{n_1, P_e^{SR}}(k) 
\end{equation}
Correspondingly, the probability that $D$ is able to decode after $M$ transmissions can be calculated as:
\begin{eqnarray}
\label{eqn:p3}
&& \hspace{-36pt} p^{(3)}(M) = g_{n_3, P_e^{SD}}(M)\nonumber \\
&& \hspace{-23pt}  \sum_{j\geq M} g_{n_1, P_e^{SR}}(j) \sum_{k \geq M} g_{n_2, P_e^{SR}}(k) + (1 \!-\!P_e^{eq}) \sum_{j<M} \! \tilde{q}(j) \! \nonumber \\
&& \hspace{-10pt}  \sum_{s=0}^{j} \!\!\! \sum_{t=0}^{M\!-\!j\!-\!1} \!\!\! \beta_{j, P_e^{SD}}(s) \beta_{M\!-\!j\!-\!1, P_e^{eq}}(t) f_{n_3}(s\!+\!t\!+\!1)
\end{eqnarray}
where $P_e^{eq} \! = \! 1\!-\!(P_e^{SD}(1\!-\!P_e^{RD}))$ is the equivalent erasure probability of the current message block.

Initially, all $n_1$, $n_2$, and $n_3$ are equal to zero. Whenever any of the $R$'s is able to decode before $D$ (let $R_1$ be the successful relay), there will be some network coded packets transmitted, therefore some information about the next message block will be available at $D$ as well as $R_2$. Let $j$ and $M$ ($M \! > \! j$) be the time at which $R_1$ and $D$ decodes the current block respectively. Since $D$ can decode the current message block, it can remove its contribution from the network coded packets; therefore the content of buffer 2 and 3 can be combined to provide the information about the next message block.

During the $M\!-\!j$ transmissions in phase two, $D$ would have filled buffer 2 or 3 in $1\!-\!P_e^{SD}$ of the time. Number of packets $\tilde{n_3}$ for the next message block is then binomially distributed according to $\beta_{M-j, P_e^{SD}}(\tilde{n_3})$. Meanwhile, there are two scenarios to be considered for $R_2$. When it is also able to decode the current block, which occurs with probability:
\begin{equation*}
\gamma = \sum_{s=0}^{j} \sum_{t=0}^{M-j} \beta_{j, P_e^{SR}}(s) \beta_{M-j, P_e^{eq2}}(t) f_{n_2}(s+t)
\end{equation*}
where $P_e^{eq2} \! = \! 1\!-\!(P_e^{SR}(1\!-\!P_e^{RR}))$ is the equivalent erasure probability, the content of buffer 2 and 3 can be combined. Therefore, the equivalent erasure probability for the next message block at $R_2$ during $M\!-\!j$ phase two transmissions is $P_e^{SR}$. Correspondingly, $\tilde{n_2}$ for the next message block is binomially distributed according to $\beta_{M-j, P_e^{SR}}(\tilde{n_2})$.

For the case when $R_2$ is not able to decode the current message block (which happens with probability $1\!-\!\gamma$), only the packets stored in buffer 3 can be used to provide information about the next message block. In this scenario, since only $(1-P_e^{SR})(1-P_e^{RR}) = 1-P_e^{eq3}$ of the transmissions go to buffer 3, the number of packets $\tilde{n_2}$ for the next message block is binomially distributed according to $\beta_{M-j, P_e^{eq3}}(\tilde{n_2})$.

As far as $R_1$ is concerned, due to half duplex constraint, it will not be receiving any packets about the next message block, therefore $\tilde{n_1}=0$ for the next block transmission. Using $\tilde{n_1}$, $\tilde{n_2}$, and $\tilde{n_3}$, transmission will then continue with phase one transmission again until all message blocks are sent.

\section{Application to Wireless Channels}
\label{Sec:AppsToWireless}
In wireless channel, instead of erasure, fading and AWGN (Additive White Gaussian Noise) contribute to channel impairments. The channel model between $S$ and $D$ in this case can be expressed using its discrete-time equivalent model:
\begin{equation}
y_D = h_{SD} x_S + \eta
\end{equation}
where $h_{SD}$ is the Rayleigh flat fading channel coefficient, distributed according to $\mathcal{CN}(0, 1/\lambda_{SD})$ ($\lambda_{SD}$ is the path loss parameter). $\eta$ is the AWGN noise, normally distributed with variance $\sigma_n^2$. $x_S$ and $y_D$ are the transmitted and received signal respectively. For other node pairs, the subscript notation can be changed accordingly. Two approaches to extend the earlier results into wireless channel scenario are given as follows.

\subsection{Approach 1: Convert wireless channel into erasure channel}
Here, the idea is to use a Gaussian codebook\footnote{The use of Gaussian codebook is to simplify the exposition on how the proposed scheme works in wireless channel. In practical scenario, good channel code with reasonable encoding/decoding complexity needs to be used.}, and map every possible $2^{m+\mu}$ encoded packet into one of the valid codewords. The encoding and decoding process can be explained using standard random coding argument as follows:

\textbf{\textit{Encoding:}} Generate $2^{nR}$ length-$n$ codewords, comprising of typical sequences drawn according to $\mathcal{N}(0,\sigma_{tx}^2)$. Here, $\sigma_{tx}^2$ is the available power of the transmitting nodes, which are assumed to be equal for all nodes. Then, a one-to-one mapping is performed to associate each one of the $2^{m+\mu}$ encoded packets to a valid codeword. Transmission of a particular encoded packet is done by transmitting the corresponding codeword to the channel.

\textbf{\textit{Decoding:}} Perform coherent detection (equivalent to division by channel fading $h$). Decoding is then performed by searching in the Gaussian codebook, the length-$n$ sequence which is jointly typical with the equalised received sequence. The desired encoded packet can then be obtained by inverse mapping, using the same table as that in the encoder.

Following Shannon's theorem of reliable communication, since $x$ is drawn from Gaussian codebook and $\eta$ is normally distributed (independent to $x$), the capacity of the above channel is $C = 0.5 \text{log}_2(1 + |h_{SD}|^2 \sigma_{tx}^2/\sigma_{n}^2)$. As long as $R < C$, any codeword can be decoded with vanishing error probability as the codeword length $n$ grows. This forms the first condition for reliable packet transmission.

The second condition is related to the mapping table used to associate each encoded packet with a valid codeword. Since the mapping is required to be one-to-one, it is necessary that $nR \geq m+\mu$. Combining with the first condition, the constraint for successful packet transmission can be written as:
\begin{equation}
0.5 \: \text{log}_2(1 + |h_{SD}|^2 \sigma_{tx}^2/\sigma_{n}^2) > (m+\mu)/n
\end{equation}
Whenever this constraint is violated, there is no rate $R$ which satisfies both of the above conditions, hence it is not guaranteed that the packet can be decoded without error.

Considering the non-decodable packet as erased, the channel can be treated as erasure channel with erasure probability:
\begin{eqnarray}
P_e^{SD} & \hspace{-11pt} = &\hspace{-8pt} \text{Pr}\left[ 0.5 \: \text{log}_2(1 + |h_{SD}|^2 \sigma_{tx}^2/\sigma_{n}^2) \leq (m+\mu)/n \right] \nonumber \\
&\hspace{-11pt} =&\hspace{-8pt} 1 - \text{exp}\left(- \lambda_{SD} \frac{2^{2(m+\mu)/n}-1}{\sigma_{tx}^2/\sigma_{n}^2} \right)
\end{eqnarray}
With the above strategy, each point-to-point channel has been converted to its equivalent erasure channel. Therefore, the earlier analysis on phase one transmission can be readily applied. For the phase two transmission, however, both $S$ and successful $R$ are transmitting simultaneously; therefore a form of multiuser detection is required at $D$. Since Gaussian codebook is used, decoding can be performed successively \cite{Multiple:Gamal}. First, the codeword of one source is decoded. Then it is subtracted out from the received signal, followed by decoding of the other source. The erasure probabilities in this case is determined by the decoding order used. When the channel from $R$ to $D$ is stronger, i.e. $|h_{RD}|^2 \geq |h_{SD}|^2$ (which happens with probability $\lambda_{SD}/(\lambda_{SD}+\lambda_{RD})$), the packet from $R$ is decoded first. The corresponding erasure probability is:
\begin{eqnarray}
\hspace{-1pt}P_e^{RD(a)} &\hspace{-6pt}=&\hspace{-6pt} \text{Pr}\left[ 0.5 \text{log}_2(1 + \frac{|h_{RD}|^2 \sigma_{tx}^2}{|h_{SD}|^2 \sigma_{tx}^2 + \sigma_{n}^2}) \leq (m+\mu)/n \right] \nonumber \\
&\hspace{-6pt}=&\hspace{-6pt} 1 - \frac{\lambda_{SD}/\chi}{\lambda_{RD} + \lambda_{SD}/\chi} \text{exp}\left(- \lambda_{RD} \: \chi \: \sigma_n^2/\sigma_{tx}^2 \right) \\
\hspace{-1pt}P_e^{SD(a)} &\hspace{-6pt}=&\hspace{-6pt} \text{Pr}\left[ 0.5 \text{log}_2(1 + |h_{SD}|^2 \sigma_{tx}^2/\sigma_{n}^2) \leq (m+\mu)/n \right] \nonumber \\
&\hspace{-6pt}=&\hspace{-6pt} 1 - \text{exp}\left(- \lambda_{SD} \: \chi \: \sigma_{n}^2/\sigma_{tx}^2 \right)
\end{eqnarray}
where $\chi = 2^{2(m + \mu)/n}-1$. On the other hand, when $|h_{RD}|^2 < |h_{SD}|^2$, packet from $S$ is decoded first, and correspondingly:
\begin{eqnarray}
\hspace{-24pt}P_e^{SD(b)} &\hspace{-6pt}=&\hspace{-6pt} 1 - \frac{\lambda_{RD}/\chi}{\lambda_{SD} + \lambda_{RD}/\chi} \text{exp}\left(- \lambda_{SD} \: \chi \: \sigma_n^2/\sigma_{tx}^2 \right) \\
\hspace{-24pt}P_e^{RD(b)} &\hspace{-6pt}=&\hspace{-6pt} 1 - \text{exp}\left(- \lambda_{RD} \: \chi \: \sigma_{n}^2/\sigma_{tx}^2 \right)
\end{eqnarray}
It is apparent that there is a dependency to the above erasure probability. Namely, when packet from $R$ is decoded first, erasure probability for packet from $S$ is equal to $P_e^{SD(a)}$ only when packet from $R$ can be successfully decoded (with probability $1-P_e^{RD(a)}$), otherwise it will not be decodable. It should also be noted that when both packets from $S$ and $R$ can be decoded, apart from the encoded message of the next block, the encoded message of the current block will also be decodable. The buffer assignment table as well as the equivalent probabilities are given in Table \ref{tab:Buffer2}. The overall performance can then be analysed using the same approach as that in ideal erasure channel.
\begin{table}[t]
	\centering
		\caption{Buffer assignment table at destination during phase two.}
		\label{tab:Buffer2}
		\begin{tabular}{c}
					\includegraphics[width=228pt]{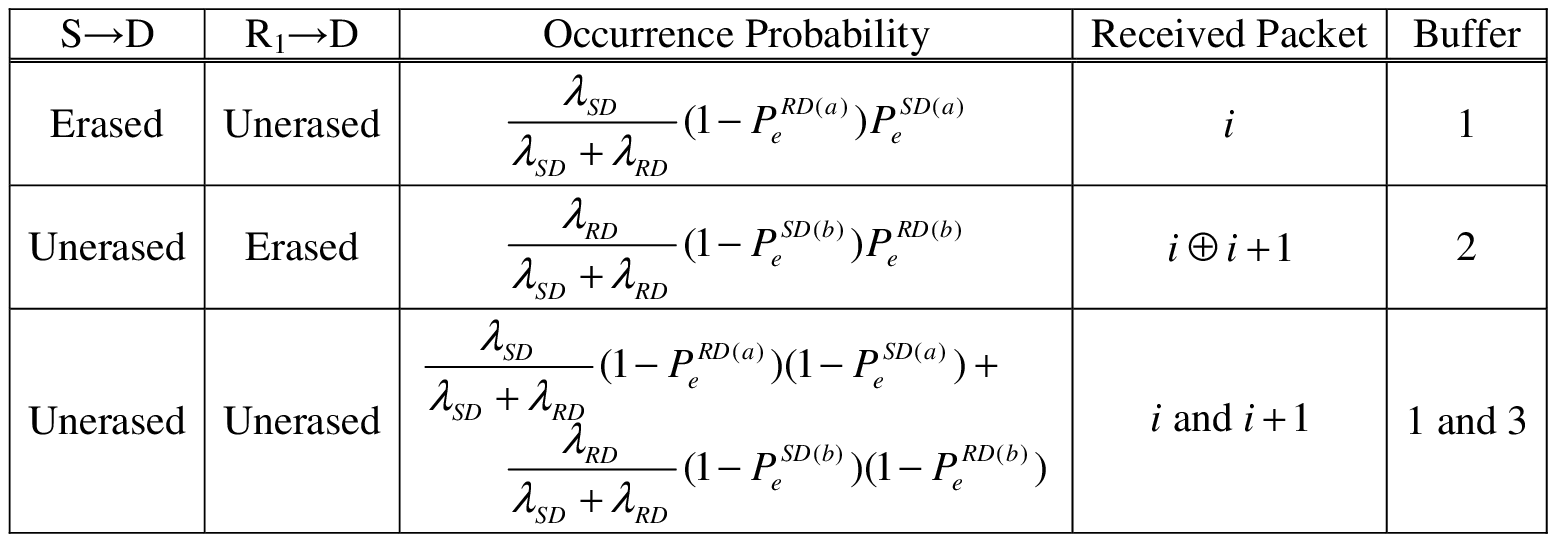}
		\end{tabular}
\end{table}

\begin{figure}[t]
	\centering
		\includegraphics[width=200pt]{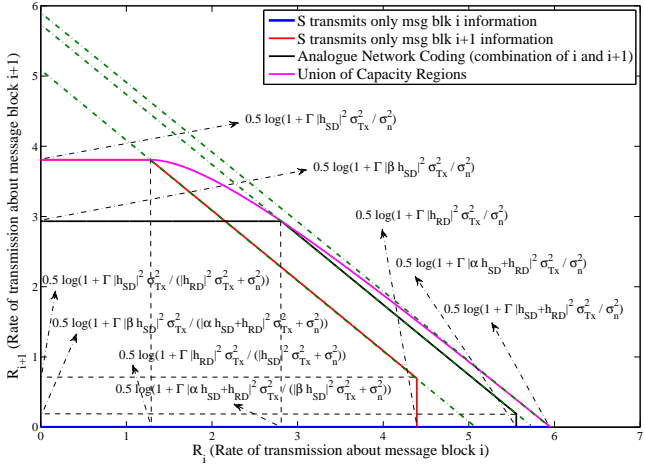}
	\caption{Capacity region of network coded cooperative scheme}
	\label{fig:CapacityRegion}
\end{figure}

\subsection{Approach 2: Assume that fountain code achieves capacity}
In this approach, we focus more on the amount of information to be sent across to destination, which is equal to $K m$ bits in each block. Denoting the capacity between $S$ and $D$ as $C_{SD} = 0.5 \text{log}_2\left(1 + \Gamma |h_{SD}|^2 \sigma_{tx}^2/\sigma_n^2 \right)$, the duration of phase one transmission can be calculated as $T_{s1}=K m/C_{SR}$. The constant $\Gamma$ is the SNR-gap used to cater for the imperfection of the actual modulation used as compared to Gaussian codewords \cite{MMSE:Cioffi}-\cite{SNR:Ana}. During this $T_{s1}$ period, $D$ also received $C_{SD} T_{s1}$ bits of information. Therefore, the remaining information about the current block that $D$ requires during phase two is $K m - (C_{SD} T_{s1})$ bits.

Following the proposed scheme, both $S$ and $R_1$ transmit to $D$ in phase two. Here, there are several possibilities. In one extreme, $S$ can transmit information about the same message block (denoted as $x_i$). The received signal at $D$ is then:
\begin{equation}
y = h_{SD} x_i + h_{RD} x_i + \eta
\end{equation}
After coherent detection is performed with respect to the combined channel, the resulting capacity can be expressed as:
\begin{equation}
C = 0.5 \text{log}_2\left(1 + \Gamma |h_{SD} + h_{RD}|^2 \sigma_{tx}^2/\sigma_n^2 \right)
\end{equation}
With this approach, the duration of phase two transmission is minimised. However, no information about the next message block is sent, hence similar problem arises. In the other extreme, $S$ can transmit information about the next message block. The received signal model at the destination is then:
\begin{equation}
y = h_{SD} x_{i+1} + h_{RD} x_i + \eta
\end{equation}
It is observed that the above model is a multiple access channel model. The capacity region of this channel is shown by the red line in Figure \ref{fig:CapacityRegion}. As compared to the previous case ($S$ transmits only block $i$ information, whose capacity region is reflected as the blue curve in Figure \ref{fig:CapacityRegion}), the rate of transmission about the current block is smaller. Therefore the duration of phase two is also longer. On the other hand, some information about the next message block will also get transmitted to both $R$ and $D$, which will expedite the transmission of subsequent message blocks. However, the advantage of having some information about the next message block versus the penalty due to longer phase two duration needs to be studied.

Having considered the two extreme cases, it is natural to analyse the condition between those two extremes. In this case, $S$ transmits a combination of the message from current block $i$ and the next block $i+1$, resulting in the following: 
\begin{eqnarray}
y &=& h_{SD} (\alpha x_i + \beta x_{i+1}) + h_{RD} x_i + \eta \nonumber \\
&=& (\alpha h_{SD} + h_{RD}) x_i + \beta h_{SD} x_{i+1} + \eta
\end{eqnarray}
This scheme is also known as analogue network coding or superposition coding \cite{Cooperative:Bergmans}. To satisfy the transmit power limitation, additional constraint $\alpha^2 + \beta^2 = 1$ is imposed. It is apparent that the above channel is another multiple access channel, with channel gains $(\alpha h_{SD} + h_{RD})$ and $(\beta h_{SD})$, which are used to transmit information about current and the next block respectively. The capacity region of this scheme is shown by the black curve in Figure \ref{fig:CapacityRegion}.

Several observations can be made about the new capacity region. Firstly, as the value of $\alpha$ is increased, the rate for transmitting the next message block $i+1$ decreases. Secondly, as $\alpha$ increases, the rate of transmitting the current message block $i$ is increased, reducing the time required in phase two transmission. Lastly, another interesting observation is that the sum rate capacity (indicated by the doted green line in Figure \ref{fig:CapacityRegion}) increases with parameter $\alpha$, which suggests that larger $\alpha$ is better. However, $\alpha$ should not be too large, as otherwise no information about the next block can be delivered.

Taking the union of all capacity regions as parameter $\alpha$ is varied, the entire capacity region of network coded transmission scheme can be found (depicted as the pink curve in the Figure \ref{fig:CapacityRegion})). Optimal operating point lies in the boundary line of this region. However, further investigation is required to find the best trade-off point. One good operating point is to set the transmission rate for the current message block to be equal to the rate when only $R$ transmits during phase two. In this way, the duration of phase two transmission is left unchanged. Meanwhile, by operating at the boundary line of the capacity region, the amount of information about the next message block at the destination (and receiving relay) is maximised.

\begin{figure}[t]
	\centering
		\includegraphics[width=162pt]{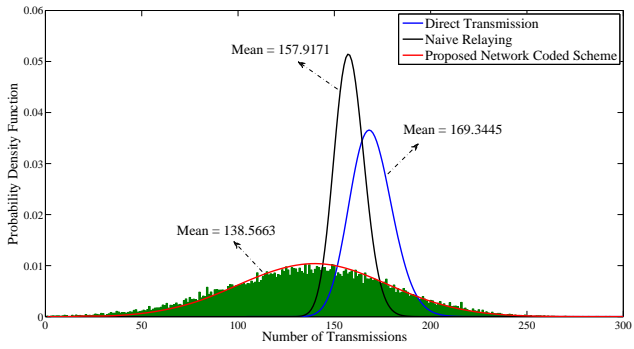}
	\caption{Histogram and the corresponding PDF of the required number of transmissions under different schemes}
	\label{fig:Histogram}
\end{figure}

\section{Numerical Analysis}
\label{Sec:Numerical}
In this section, the performance of various schemes in terms of required number of transmissions are analysed. Due to the nature of fountain code, no analysis on bit error rate is given, as transmission is allowed to continue until successful decoding is achieved at destination.  A system with single destination and two relay nodes is considered. The erasure probability of the direct link is set to $P_e^{SD} \hspace{-3pt}=\hspace{-3pt} 0.4$, while that of the indirect links is set to $P_e^{SR} \hspace{-3pt}=\hspace{-3pt} P_e^{RD} \hspace{-3pt}=\hspace{-3pt} P_e^{RR} \hspace{-3pt}=\hspace{-3pt} 0.2$. The number of packets in each message block is $K\hspace{-3pt}=\hspace{-3pt}100$, and transmissions are normalised with respect to packet size $m$. Figure \ref{fig:Histogram} shows the histogram as well as the corresponding PDF (calculated using equation (\ref{eqn:p3})) of the required number of transmissions using the proposed network coded scheme. The PDF of the number of transmissions under direct transmission and naive relaying schemes (calculated according to equation (\ref{eqn:p1}) and (\ref{eqn:p2}) respectively) are also shown for comparison.

It is observed from the figure that the proposed scheme outperforms both direct transmission and naive relaying, which is reflected by the smaller mean in the required transmissions. It is to be noted, however, that the variance is larger; which signifies that it is more suitable for non delay-constrained applications. Another interesting observation is that unlike the other two schemes, there is a positive probability that successful decoding is possible in less than $K$ transmissions. This is attributed from the extra information in the form of encoded packets, which are received during the previous block.

\begin{figure}[t]
	\centering
		\includegraphics[width=162pt]{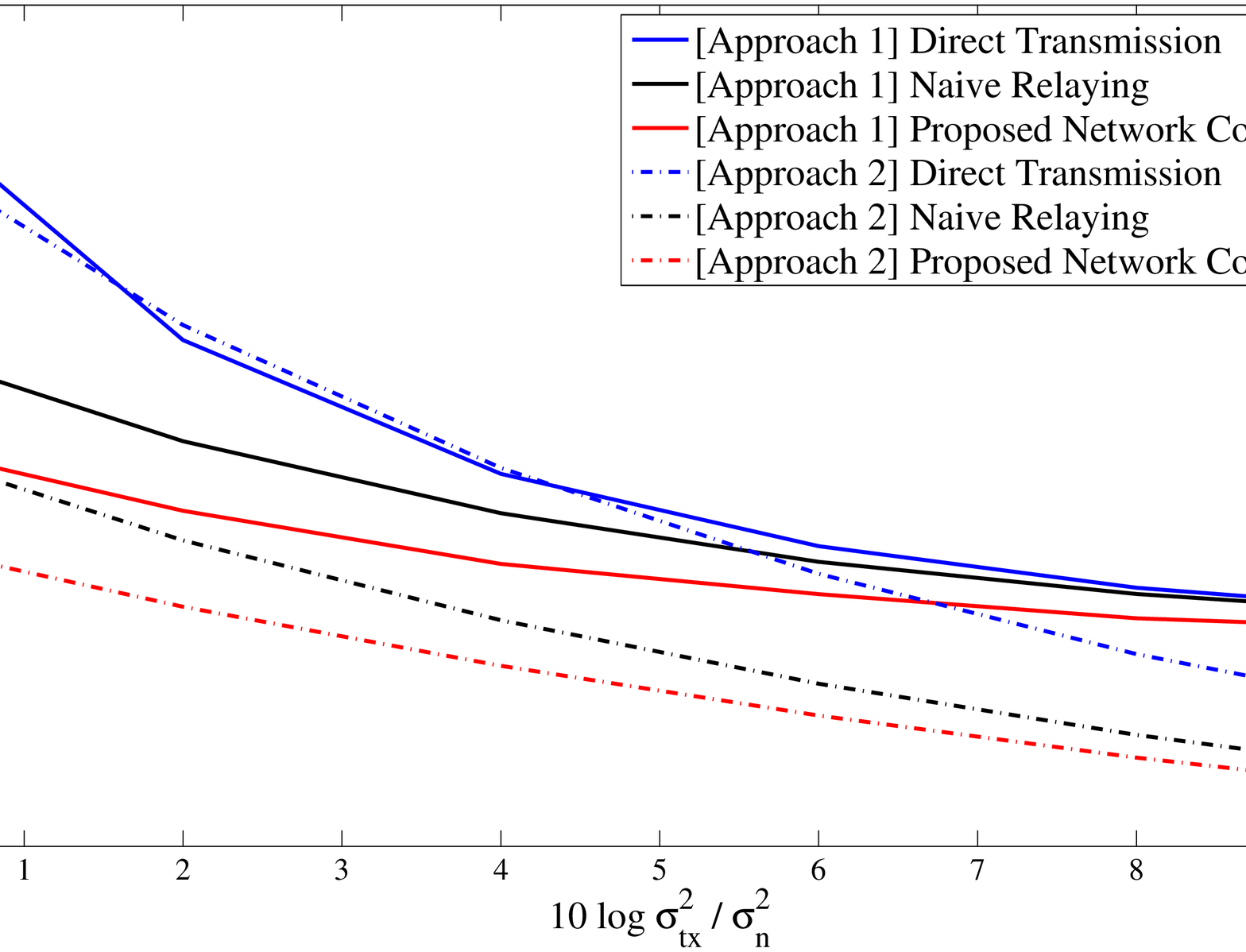}
	\caption{Average number of transmissions for successful decoding over wireless channel as a function of SNR}
	\label{fig:MeanTransmissions}
\end{figure}

In Figure \ref{fig:MeanTransmissions}, the average required number of transmissions over wireless channel is depicted for the three schemes under consideration. Here, the path loss parameter is related to distance according to $\lambda \propto d^\alpha$, where path loss exponent $\alpha$ is set to 3, and the source-destination distance ($d_{SD}$) is set to $20$ meters. Meanwhile, the distance of the relay with respect to both source and destination is set to $d_{SR} \hspace{-1pt} = \hspace{-1pt} d_{RD} \hspace{-1pt} = \hspace{-1pt} 10.3$ meters, and the inter-relay distance is $d_{RR} \hspace{-1pt} = \hspace{-1pt} 5$ meters. From Figure \ref{fig:MeanTransmissions}, it is observed that the schemes exhibit similar behaviour in wireless channel (for both approaches), whereby the proposed network coded scheme outperforms the other two schemes. The difference is more prominent in low SNR regime. In high SNR regime, the effective erasure probability is very small, hence the performance of the three schemes are comparable.

\section{Conclusion}
\label{Sec:Conclusion}
In this paper, a novel transmission scheme of fountain code over cooperative relay networks with multiple relay nodes is proposed. The scheme combines network coding concept to allow source node transmitting partial information about the next message block during the current block transmission. This extra information will expedite the transmission of subsequent blocks, thereby improving the overall throughput. Two possible approaches to implement the scheme into wireless channel are discussed. It is then shown via numerical analysis that in both approaches, the proposed scheme is able to outperform the naive relaying scheme, especially in low SNR regime. Future work includes a search for optimal operating point as well as a development of robust scheme, which could adapt the relaying strategy according to channel conditions.

\end{document}